\documentclass[twocolumn,floatfix,superscriptaddress,a4paper,showkeys,nofootinbib,prl]{revtex4-2}
\usepackage{epsfig}
\usepackage{latexsym}
\usepackage{xspace}
\usepackage[colorlinks=true,linktocpage=true,linkcolor=blue,citecolor=blue,allcolors=blue]{hyperref}
\usepackage{url}
\usepackage[utf8]{inputenc}
\usepackage{indentfirst}
\usepackage{enumerate}
\usepackage{color}
\usepackage{comment}

\usepackage[caption=false,position=top]{subfig}

\usepackage{amsmath}
\usepackage{amssymb}

\usepackage[english]{babel}
\usepackage{url}

\AtBeginDocument{}

\newcommand{\mean}[1]{\langle #1 \rangle}

\newcommand{\eq}[1]{\begin{align} #1 \end{align}}

\begin{document}
\title{
Sensitivity of transverse momentum correlations to early-stage and thermal fluctuations
}

\author{Oleh Savchuk}\thanks{Corresponding author}
\email{savchuk@frib.msu.edu} 
\affiliation{Facility for Rare Isotope Beams, Michigan State University, East Lansing, MI 48824 USA}
\affiliation{Bogolyubov Institute for Theoretical Physics, 03680 Kyiv, Ukraine}

\begin{abstract}

Transverse momentum correlations were recently measured by the ALICE collaboration at the LHC \cite{ALICE-Ridge}. A long-range structure in terms of relative pseudorapidity of particle pairs is observed. This may imply some signal of the initial state owing to the shear spread of the correlation. However, the fluctuations inside a thermally equilibrated medium have to be taken into account, serving as motivation for this letter. Using lattice Quantum Chromodynamics (lQCD) constraints, we predicted the development and spread of balancing correlations caused by energy-momentum conservation. Simultaneously, we propagated the initial correlation using hydrodynamics to estimate its effects. Our findings suggest that the resulting correlation, known as ``the ridge``, is sensitive to both fluctuations in the equilibrated medium and the pre-equilibrium stage. This can provide important insight into the early stages of the collision.

\end{abstract}

\maketitle

\paragraph{\bf Introduction.}
The structure of the Quantum Chromodynamics (QCD) phase diagram 
is one of the major open questions in high energy physics, and has attracted great theoretical and experimental effort ~\cite{Busza:2018rrf,Bzdak:2019pkr}. Fluctuations and correlations of baryonic, electric and strange charges have consistently held a significant role in characterizing the bulk properties of matter, particularly in proximity to phase transitions or critical points~\cite{Steinheimer:2012gc,Savchuk:2022msa,Vovchenko:2020tsr, Sorensen:2022odd}. Fluctuations of energy and momentum also probe the finest features of the equation of state and are related to the microscopic structure of the matter. On the other hand, it is believed that fluctuations of energy and momentum in Heavy-Ion Collisions are influenced by the hard processes that take place before the equilibrium medium is created. Thus, understanding the connection between fluctuations of energy or momentum and the equation of state is a major challenge. 

Studies of soft particles (mean transverse momentum $\mean{p_t}<2$~GeV) revealed a complex structure of correlations in relative pseudorapidity, $\Delta \eta$, and  azimuthal angle, $\Delta \phi$, known as ``the ridge``. It has been suggested that the ridge is caused by early stage correlations that extend over several units of rapidity~\cite{PhysRevC.79.051902,Dumitru:2008wn,Kharzeev:2000ph}. The interaction of jets with the bulk of the matter can contribute as well~\cite{PhysRevC.76.047901,Moschelli_2009}. The soft ``ridge`` has been linked to viscous relaxation in the medium, predicting a dependence on viscosity of the quark-gluon plasma~\cite{Gavin:2008ta, PhysRevLett.97.162302}. The detailed simulation of the hydrodynamic response to thermal fluctuations and viscous relaxation in the framework of~\cite{first_paper} is the primary focus of this letter. 

Recent measurements of flow coefficients and hydrodynamical predictions largely focus on the precise determination of shear viscosity to entropy ratio $\frac{\eta_s}{s}$~\cite{PhysRevC.72.014904,Heinz_2013}.
Previously, correlations of electric, strange and baryonic charge, known as balance functions have been employed to constrain the diffusive properties of the QGP~\cite{PhysRevC.104.014906}. In this work, similar treatments are applied to correlations of energy and momentum. It is expected that results will show sensitivity to both the equation of state and the viscosity. 

Transverse momentum correlations were recently measured by the ALICE collaboration at the LHC \cite{ALICE-Ridge}. Similar measurements were performed by the STAR collaboration at RHIC \cite{STAR:2011iio}. Interpreting these measurements represents the central goal of our exploration.

\paragraph{\bf Theoretical Method.}

In the specific scenario of vanishing baryon density, we employ the formalism outlined in reference \cite{first_paper}. Assuming a (1+1)D model of Heavy-Ion Collision, with space-time rapidity $\eta$ and proper time $\tau$, the correlation in the system is divided into two parts:
\begin{eqnarray}
C_{AB}(\Delta \eta,\tau)&=&\chi_{AB}(\tau)\frac{\delta(\Delta \eta)}{\tau}+C_{AB}^{\prime}(\Delta \eta,\tau).
\end{eqnarray}
In this context, $A, B$ correspond to the energy-momentum and baryonic, electric, and strange charges, forming a 7×7 matrix. If a system is equilibrated, the strength of the local correlation is determined by the charge susceptibility, a fundamental bulk property of the matter. The second term, the balancing correlation, is spread and constrained by the time elapsed since the correlations were seeded.

With a set of Green's functions, denoted as \(G_{AB}(\eta_f-\eta_i, \tau_f,\tau_i)\), which describe the response in \(B\) at \(\tau=\tau_f, \eta=\eta_f\) to the perturbation in \(A\) at \(\tau=\tau_i, \eta=\eta_i\), the correlation at \(\Delta \eta \neq 0\) is expressed as follows:

\eq{\label{cab}
&C_{AB}^{\prime}(\Delta \eta = \eta_2-\eta_1,\tau_f)=\int\limits_{\tau_0}^{\tau_f} d\tau_j \tau_j \int d\eta_j G_{A A^\prime}(\eta_1-\eta_j,\tau_f,\tau_j)\nonumber\\
&G_{B B^\prime}(\eta_2-\eta_j,\tau_f,\tau_j)S_{A^\prime B^\prime}(\tau_j),
}
where the limits in time integration correspond to the initial time of evolution \(\tau_0\) and the freeze-out time \(\tau_f\) and $S(\tau_j)$ is the source:
\eq{
S_{AB}(\tau)=(\partial_{\mu}u^{\mu})\chi_{AB}=\left(\frac{1}{\tau}+\partial_{\tau}\right)\chi_{AB}(\tau).
}
However, Equation (\ref{cab}) does not account for the initial correlation that should naturally exist in the equilibrated hydrodynamic medium right from the start. To include this initial correlation, we introduce the term:

\eq{\label{cab_0}
&C_{AB}^{0\prime}(\Delta \eta = \eta_2-\eta_1,\tau_f)= \tau_0^2\int d\eta_{j_1}d\eta_{j_2} G_{A A^\prime}(\eta_1-\eta_{j_1},\tau_f,\tau_0)\nonumber\\
&G_{B B^\prime}(\eta_2-\eta_{j_1},\tau_f,\tau_0)C_{A^\prime B^\prime}^{0\prime}(\eta_{j_2}-\eta_{j_1},\tau_0),
}

Here, based on the conservation of charge, we acknowledge that \(\int d\eta_{j_2}C^0_{AB}(\eta_{j_1}-\eta_{j_2},\tau_0)=-\chi_{AB}(\tau_0)/\tau_0\) to satisfy sum rules related to conservation laws~\cite{first_paper}. However, the precise shape of the function remains unknown. For the purposes of our study, we make the assumption that the initial correlation corresponds to a Gaussian with a width \(\sigma_{AB}\):

\eq{\label{cab-0}
C_{AB}^0(\eta_{j_2}-\eta_{j_1},\tau_0)=-\frac{\chi_{AB}(\tau_0)}{\sqrt{2\pi \sigma_{AB}^2}\tau_0}\exp^{-\frac{(\eta_{j_2}-\eta_{j_1})^2}{2\sigma_{AB}^2}}.
}

Determining the exact shape of the correlation for the initial state of the hydrodynamic simulation would necessitate a comprehensive transport study. Although this study in itself holds significance, our conclusions will primarily focus on the part of the correlation developed during the hydrodynamic stage, as outlined in Eq.(\ref{cab}). At the same time, viscous spread of the initial correlation will make any structure on the scale of \(\Delta \eta < 1\) smeared out in Eq.(\ref{cab_0}).

Green`s functions can be obtained from relativistic hydrodynamics~\cite{Romatschke:2010}. For the Bjorken solution, $\rho(t,\vec{r}),\varepsilon(t,\vec{r})$ and $u^{\mu}(t,\vec{r})$ (electric, strange, baryon and energy densities and four-velocity) can be expanded into a series, with respect to the perturbations $\delta \rho,\delta \varepsilon ,\delta u^{\mu}$.
The perturbation is described by a set of linear equations:
\eq{
&\partial_{\tau}\delta\varepsilon = -\frac{1}{\tau}\delta\left(\varepsilon+P\right) - \partial_{\eta}\frac{\varepsilon+P-\frac{8\eta_s}{3\tau}}{\tau}\delta u^{\eta}+j_{\varepsilon}, \\
&\partial_{\tau}\left(\varepsilon+P-\frac{4\eta_s}{3\tau}\right)\delta u^{\eta} = -\frac{2}{\tau}\left(\varepsilon+P-\frac{4\eta_s}{3\tau}\right)\delta u^{\eta} \nonumber \\
&- \partial_{\varepsilon} P\frac{\partial_{\eta}\delta\varepsilon}{\tau} + \frac{4\eta_s\partial_{\eta}^2\delta u^\eta}{3\tau^2}+j_{P_\eta}, \\
&\partial_{\tau}\delta\rho = -\frac{\delta \rho}{\tau} + \frac{D}{\tau^2}\partial_{\eta}^2\delta\rho+j_{\rho}, \\
&\partial_{\tau}\left(\varepsilon+P-\frac{2\eta_s}{3\tau}\right)\delta u^{x,y} = -\frac{1}{\tau}\left(\varepsilon+P-\frac{2\eta_s}{3\tau}\right)\delta u^{x,y}\nonumber \\
&+\frac{\eta_s\partial_{\eta}^2\delta u^{x,y}}{\tau^2}+j_{P_{x,y}},
}
where $\delta u^{\eta}=\frac{\delta u^z}{u^0}$. For a Green`s function $j_A=\delta_{A}(\tau-\tau_j)\delta(\eta-\eta_j)/\tau$ corresponds to a point source. The connection between velocity and local momentum is:
\eq{
\delta P_{\eta}&=\left(\varepsilon+P-\frac{4\eta_s}{3\tau}\right)\delta u^{\eta},\\
\delta P_{x,y}&=\left(\varepsilon+P-\frac{2\eta_s}{3\tau}\right)\delta u^{x,y}.
}

\begin{figure}[h!]
\includegraphics[width=.49\textwidth]{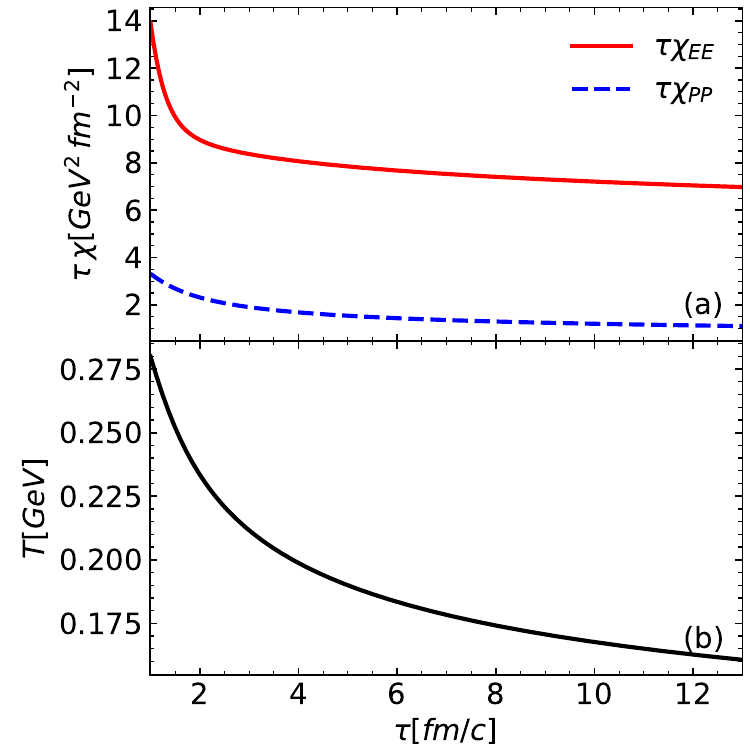}
\caption{\label{Sources} Susceptibility that creates energy-momentum correlations $\chi_{PP},\,\chi_{EE}$ (a) and temperature $T$ (b) as functions of proper time $\tau$ along the Bjorken solution with lQCD equation of state.
}
\end{figure}
The equation of state plays a pivotal role in defining the hydrodynamic evolution of the system and determines the susceptibilities essential for calculating correlations of conserved charges. For our equation of state, we utilized the results from the HotQCD collaboration \cite{Bazavov_2014} for the pressure and energy density as a function of temperature, and \cite{Bazavov_2012} for susceptibilities of conserved charges.

In the topmost plot (a) of Figure \ref{Sources}, $\chi_{EE}$ and $\chi_{PP}=\chi_{P_x P_x}=\chi_{P_y P_y}=\chi_{P_\eta P_\eta}$ are shown as monotonic functions of $\tau$. Their most rapid rate of change corresponds to the early stages of the collision, indicating that the majority of balancing correlation is generated during this period, as per Eq.(\ref{cab}). Notably, $\frac{\chi_{PP}(1)}{13\chi_{PP}(13)}\approx 4$, suggesting that the correlation created during the hydrodynamic stage should be comparable to the correlation from the pre-equilibrium stage. A similar trend is observed for $\frac{\chi_{EE}(1)}{13\chi_{EE}(13)}\approx 2$.
Panel (b) illustrates the temperature decrease resulting from expansion in the longitudinal frame. As our model assumes $\mu_B,\mu_S,\mu_Q=0$, all other thermodynamic quantities, such as $P, \varepsilon, s$ can be calculated at specific temperature values.

Given that the equation of state and susceptibilities are well constrained by lattice calculations, our focus turns to the transport coefficients, specifically viscosity ($\eta$) and diffusion ($D$). Balance function studies~\cite{PhysRevC.104.014906} provide estimates for the diffusion constant, with $D=\frac{T}{2\pi}$ and $\sigma_{AB}=0.5$, where $A\, , B$ can be baryonic, electric, and strange charges. This leaves us with two remaining parameters: $\eta_s/s$ and $\sigma_0^\eta=\sigma_{AB}$, where $A\, , B$ can be $E,P_x,P_y,P_{\eta}$, i.e. energy and momentum. We adopt $\frac{\eta_s}{s}=\frac{1}{2\pi}$ as a reasonable estimate and consider $\sigma_0^\eta={2,4,8}$.
This choice motivates by different scales of interactions. Thermal and equilibrium one usually short range, on the other hand flux tubes and minijets can spread several units of rapidity.

\begin{figure*}[t!]
\includegraphics[width=.99\textwidth]{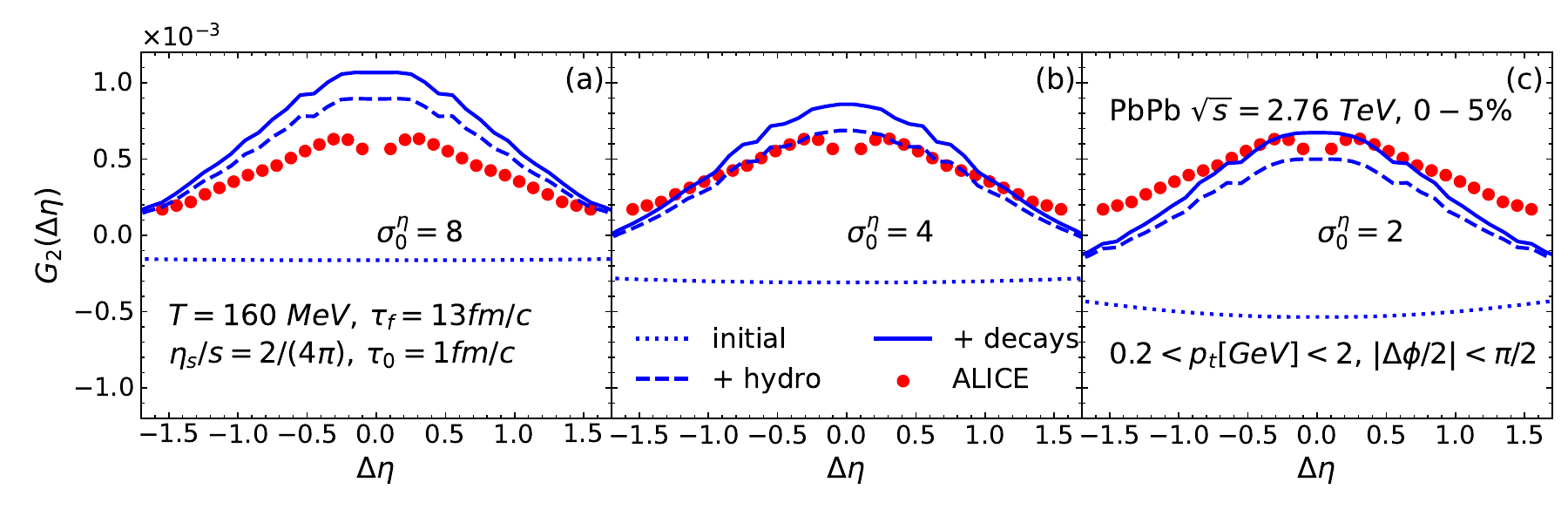}
\caption{\label{corr-out} Comparison of $G_2(\Delta \eta)$ calculations to measurements of the ALICE collaboration~\cite{ALICE-Ridge}. Freeze-out parameters were inspired by~\cite{Huovinen:2016xxq} and $\eta_s/s$ was set to $1/(2\pi)$. Panels (a), (b) and (c) display results for three separate widths, $\sigma_0^\eta$, of the preequilibrium correlation. The preequilibrium correlations, dotted lines, are added to the hydrodynamic correlations to get the dashed lines, which after the effects of decays are added yields the net correlation, denoted by solid lines. Because the preequilibrium correlation must integrate to a fixed value, which is negative, smaller values of $\sigma_0^\eta$ lower the overall correlation for small $\Delta\eta$. For $\sigma_0^\eta=4$ the net correlation comes closest to the ALICE measurement. 
}
\end{figure*}
\paragraph{\bf Transverse Momentum Correlations.}
The ALICE Colllaboration measured the quantity $G_2$,
\eq{
&G_2(\eta_1,\eta_2,\phi_1,\phi_2)= \\ &\frac{1}{\mean{p_{t,1}}\mean{p_{t,2}}}\Big{[}\frac{\int dp_{t,1} dp_{t,2} p_{t,1}p_{t,2} \rho_2(\vec{p}_1,\vec{p}_2)}{\int dp_{t,1}  \rho_1 (\vec{p}_1)\int dp_{t,2}  \rho_1(\vec{p}_2)}\nonumber \\&-\mean{p_{t,1}(\eta_1,\phi_1)}\mean{p_{t,2}(\eta_2,\phi_2)}\Big{]},\nonumber} 
in PbPb collisions at $\sqrt{s}=2.76$~TeV~\cite{ALICE-Ridge}, $\rho_2(\vec{p}_1,\vec{p}_2)={d^6N_{ch}\over d\vec{p}_1 d\vec{p}_2}$, $\rho_1(\vec{p})={d^3N_{ch}\over d\vec{p}}$. The present measurement of the $G_2$ correlator is based on charged particle tracks measured with the TPC detector in the transverse momentum range $0.2 \leq p_t \leq 2.0~\mathrm{GeV}/c$  and the pseudorapidity range $|\eta| < 0.8$. This measure was integrated over $\eta_1,\phi_1$ and $\eta_2,\phi_2$, while binned on $\Delta\eta$ and with $|\Delta\phi|<\frac{\pi}{2}$.  The resulting correlation is referred to as $G_2(\Delta \eta)$. In the ALICE measurements the correlation in the  remaining azimuthal acceptance $|\Delta\phi|>\frac{\pi}{2}$ exhibits an  almost constant correlation in $\Delta\eta$. 

To compare with experimental data in relative pseudorapidity we transform correlations in space-time rapidity obtained in hydrodynamic modeling into pairs of particles (as was explained in \cite{first_paper}, see also \cite{Pratt:2017oyf,Pradeep:2022mkf}). Each pair is created assuming a lack of correlation, but each pair is then assigned a weight based on the correlations which are functions of the relative coordinates of their emission points. If the particle decays, the weight is also transferred to all its daughters. In addition to weights deriving  from the spatial correlations, a second contribution is added that accounts for the decays between daughters of the same decay. We subtracted the background estimated by ALICE from the data. The quantity $G_2$ is not intensive and scales as $\mean{N_{ch}}^{-1}$; therefore, we scaled the results for our model by the ratio of our multiplicity to that measured by ALICE so as to make a proper comparison with ALICE measurements. 

The most important shortcoming of our model is that transverse collective flow is ignored. For that reason the mean $p_t$ from our model falls significantly below that of the experiment. Because the definition of $G_2$ includes a division by two factors of average transverse momentum, the measure $G_2$ is dimensionless. Thus, the neglect of transverse flow might not be as serious as one might expect. Neglecting transverse flow still affects the results in several aspects. First, energy-energy correlations in the fluid frame can morph into momentum-momentum correlations due to a boost. And conversely, momentum-momentum correlations can drive energy-energy correlations through boosting. Because transverse momentum, $p_t$,  is not the same as $p_x$ or $p_y$, and is not expressly a conserved quantity, it is non-trivial to estimate how $p_t-p_t$ correlations might be affected by boosting the fluid elements. Secondly, the width of the correlations in relative rapidity should be narrowed by boosting. If two correlated particles, with momentum $\vec{p}_1$ and $\vec{p}_2$ are boosted transversely, their relative rapidity decreases. Finally, the lack of transverse flow affects the duration of the collision. Transverse expansion hastens the cooling and disassembly of the matter, which in turn, reduces the relative spatial spread of the correlation. Nonetheless, our treatment here is invaluable in that it provides a rough idea of how much of the the ridge might be driven by conservation effects driven by the evolution of the thermal susceptibilities. 

Each calculation displayed in Fig. \ref{corr-out} includes three contributions. The first contribution is from the pre-equilibrium correlation, which is then propagated through the hydrodynamic stage as described in Eq. (\ref{cab_0}). This contribution is characterized by the Gaussian width of the initial distribution, $\sigma_0^\eta$. Because this correlation must integrate to a known value, which is negative, smaller widths give stronger negative correlations at small $\Delta\eta$. The second contribution, referred to here as the hydrodynamic contribution, is that which traces its origin to the source function in Eq. (\ref{cab}). Finally, the third contribution is that due to the particles coming from daughters of the same decaying resonance. Fig. \ref{corr-out} displays the cummulative sums of these contributions. 
The starting and freeze-out time, temperature and transverse radius were set to $\tau_0=1\ \mathrm{fm/c}$, $\tau_f=13\ \mathrm{fm/c}$, $T_f=160\ \mathrm{MeV}$,  $R\approx10.5\ \mathrm{fm}$, as inspired by \cite{Huovinen:2016xxq}, and the sheer viscosity was set to $\eta_s=s/(2\pi)$. This choice of parameters successfully reproduces the charged particle multiplicity of the most central PbPb collisions at $\sqrt{s}=2.76\ \mathrm{TeV}$ measured by the ALICE collaboration \cite{PhysRevC.88.044910}, with $dN_{ch}/d\eta\approx1600$. 

The three panels illustrate results for three choices of the width of the initial correlation. In Panel (a) with $\sigma_0^\eta=8$, the pre-equilibrium correlation is so wide that it appears almost constant in the displayed rapidity interval that corresponds to the experimental acceptance. In this range, equilibrium correlation has a leading contribution to the correlation measure, and $G_2$ stays positive. The full correlation overshoots the data by a roughly constant factor of $1.7$. In panel (b), $\sigma_0^\eta$ is reduced to four units of rapidity, which then leads to stronger negative correlations in the rapidity range. The net correlation is lowered even further in panel (c), where the $\sigma_0^\eta$ was set to two units. The calculation using $\sigma_0^\eta=4$ came closest to the ALICE measurements.

To characterize the width of the distribution, we fit the calculation with:
\eq{
G_2(\Delta \eta)=A\exp{-\frac{\Delta \eta^2}{2\sigma_{G_2}^2}},
}
in the $-1.5<\eta<1.5$ interval where the data is available. The constant $A$ takes into account the overall normalization. In the following, we ignore possible deviations of the shape from Gaussian; however, a more involved distribution can be used. Results of the fitting are presented in Table \ref{tab:rn}. Notably, the width extracted from the root mean square of the data ($0.76$)~\cite{ALICE-Ridge} aligns with the width of our model ($0.6-0.9$). The contribution from decays appears to be moderate, at $10\%$, with the width matching the hydrodynamic part. 
\begin{table}
\begin{tabular}
{| l | c | c | c |}
\hline
\hline
 $\sigma_0^\eta$&~2~&~4~&~8~\\
\hline
        $\sigma_{G_2}$       & 0.63  & 0.76   & 0.88 \\
   \hline   
   \hline
\end{tabular}
\caption{
\label{tab:rn} 
The width of the balancing correlation $G_2(\Delta \eta)$ extracted from a gaussian function fit as a function of the spread of pre-equilibrium correlation $\sigma_0^\eta$.}
\end{table}

 The width is influenced by the choice of parameters in our model. The spread of the Green`s functions in Eqs.(\ref{cab}) and (\ref{cab_0}) increases or decreases with the viscosity and time elapsed. As the viscosity is changed in the wide range $\eta_s=(1-5)s/(4\pi)$ the width of the correlation changes by $20\%$. However the integrated strength of balancing correlation remains approximately the same. If the starting time of hydro-simulation is decreased from $1~\mathrm{fm}/c$ to $0.5~\mathrm{fm}/c$, the width of pre-equilibrium part in Eq.(\ref{cab_0}) increases. Additionally, the strength of pre-equilibrium correlation changes according to Eq. (\ref{cab-0}). We confirmed that changing $0.5~{\mathrm fm}/c \leq \tau_0 \leq 1~\mathrm{fm}/c$ does not lead to the width variation of more than a few percent. 

The predominantly positive correlation in a 3-unit range around mid-pseudorapidity suggests that $\sigma_0^\eta \sim 4$ is needed to match the data in magnitude. In all cases studied, the majority of the resulting correlation in the $|\Delta \eta| < 1.5$ range comes from the hydrodynamic contribution. This implies that the "ridge" structure derives most of its strength from the thermalized quark-gluon plasma formed in heavy-ion collision and not in the pre-equilibrium stage.

\paragraph{\bf Conclusions.}

The hydrodynamic response to locally thermalized fluctuations was utilized to estimate transverse momentum correlations as a function of pseudorapidity difference. This response ensures overall energy, charge, and momentum conservation, dependent on the transport properties of the medium. However, it's important to note that our model lacks an essential element in transverse expansion. Therefore, one should be cautious not to overinterpret the surprisingly good fit to the experimental data.
The results do indicate that the model and the $G_2$ correlation measured by the ALICE collaboration are of the same order. This matching could potentially be improved by using (3+1)D hydrodynamics and incorporating a post-hydrodynamic kinetic stage. 

We have identified the shape of the early stage as the primary source of uncertainty in modeling transverse momentum correlations. This characteristic makes transverse momentum correlations a compelling observable within the realm of initial state studies. The pre-equilibrium correlations are expected to exhibit a significant width, with $\sigma_0^\eta \sim 4$, indicating long-range correlations in rapidity. This broadening is further amplified by the collective motion and dissipative processes within the medium.

Our findings suggest that the width of the correlation is well described with the model used in the analysis. The predicted transverse momentum correlation is sensitive to the pre-equilibrium correlations and supports previous estimates for the $\eta_s/s$ ratio. The energy-energy and momentum-momentum susceptibilities used as sources of the balancing correlation provide a connection between experimentally measured correlations and the equation of state. The positive correlation observed in the data supports the idea of the soft transverse momentum-momentum correlation being caused by thermal fluctuations in the QCD medium, serving as a possible explanation for ``the ridge.``
If one were to improve the model along the lines mentioned above, i.e., including transverse expansion and better modeling the breakup stage, the hydrodynamic contribution could be confidently predicted. This would potentially enable the extraction of pre-equilibrium contributions by analyzing discrepancies between the model and the data.
\begin{acknowledgments}
The author is thankful to S. Pratt and M. Gorenstein for fruitful comments and discussions. 
Supported by the Department of Energy Office of Science through Grant No. DE-FG02-03ER41259. 
\end{acknowledgments}

\bibliography{main.bib}

\begin{thebibliography}{26}%
\makeatletter
\providecommand \@ifxundefined [1]{%
 \@ifx{#1\undefined}
}%
\providecommand \@ifnum [1]{%
 \ifnum #1\expandafter \@firstoftwo
 \else \expandafter \@secondoftwo
 \fi
}%
\providecommand \@ifx [1]{%
 \ifx #1\expandafter \@firstoftwo
 \else \expandafter \@secondoftwo
 \fi
}%
\providecommand \natexlab [1]{#1}%
\providecommand \enquote  [1]{``#1''}%
\providecommand \bibnamefont  [1]{#1}%
\providecommand \bibfnamefont [1]{#1}%
\providecommand \citenamefont [1]{#1}%
\providecommand \href@noop [0]{\@secondoftwo}%
\providecommand \href [0]{\begingroup \@sanitize@url \@href}%
\providecommand \@href[1]{\@@startlink{#1}\@@href}%
\providecommand \@@href[1]{\endgroup#1\@@endlink}%
\providecommand \@sanitize@url [0]{\catcode `\\12\catcode `\$12\catcode `\&12\catcode `\#12\catcode `\^12\catcode `\_12\catcode `\%12\relax}%
\providecommand \@@startlink[1]{}%
\providecommand \@@endlink[0]{}%
\providecommand \url  [0]{\begingroup\@sanitize@url \@url }%
\providecommand \@url [1]{\endgroup\@href {#1}{\urlprefix }}%
\providecommand \urlprefix  [0]{URL }%
\providecommand \Eprint [0]{\href }%
\providecommand \doibase [0]{https://doi.org/}%
\providecommand \selectlanguage [0]{\@gobble}%
\providecommand \bibinfo  [0]{\@secondoftwo}%
\providecommand \bibfield  [0]{\@secondoftwo}%
\providecommand \translation [1]{[#1]}%
\providecommand \BibitemOpen [0]{}%
\providecommand \bibitemStop [0]{}%
\providecommand \bibitemNoStop [0]{.\EOS\space}%
\providecommand \EOS [0]{\spacefactor3000\relax}%
\providecommand \BibitemShut  [1]{\csname bibitem#1\endcsname}%
\let\auto@bib@innerbib\@empty
\bibitem [{\citenamefont {Acharya}\ \emph {et~al.}(2020)\citenamefont {Acharya} \emph {et~al.}}]{ALICE-Ridge}%
  \BibitemOpen
  \bibfield  {author} {\bibinfo {author} {\bibfnamefont {S.}~\bibnamefont {Acharya}} \emph {et~al.} (\bibinfo {collaboration} {ALICE}),\ }\bibfield  {title} {\bibinfo {title} {{Longitudinal and azimuthal evolution of two-particle transverse momentum correlations in Pb-Pb collisions at $\sqrt{s_{\rm{NN}}}$ = 2.76 TeV}},\ }\href {https://doi.org/10.1016/j.physletb.2020.135375} {\bibfield  {journal} {\bibinfo  {journal} {Phys. Lett. B}\ }\textbf {\bibinfo {volume} {804}},\ \bibinfo {pages} {135375} (\bibinfo {year} {2020})},\ \Eprint {https://arxiv.org/abs/1910.14393} {arXiv:1910.14393 [nucl-ex]} \BibitemShut {NoStop}%
\bibitem [{\citenamefont {Busza}\ \emph {et~al.}(2018)\citenamefont {Busza}, \citenamefont {Rajagopal},\ and\ \citenamefont {van~der Schee}}]{Busza:2018rrf}%
  \BibitemOpen
  \bibfield  {author} {\bibinfo {author} {\bibfnamefont {W.}~\bibnamefont {Busza}}, \bibinfo {author} {\bibfnamefont {K.}~\bibnamefont {Rajagopal}},\ and\ \bibinfo {author} {\bibfnamefont {W.}~\bibnamefont {van~der Schee}},\ }\bibfield  {title} {\bibinfo {title} {{Heavy Ion Collisions: The Big Picture, and the Big Questions}},\ }\href {https://doi.org/10.1146/annurev-nucl-101917-020852} {\bibfield  {journal} {\bibinfo  {journal} {Ann. Rev. Nucl. Part. Sci.}\ }\textbf {\bibinfo {volume} {68}},\ \bibinfo {pages} {339} (\bibinfo {year} {2018})},\ \Eprint {https://arxiv.org/abs/1802.04801} {arXiv:1802.04801 [hep-ph]} \BibitemShut {NoStop}%
\bibitem [{\citenamefont {Bzdak}\ \emph {et~al.}(2020)\citenamefont {Bzdak}, \citenamefont {Esumi}, \citenamefont {Koch}, \citenamefont {Liao}, \citenamefont {Stephanov},\ and\ \citenamefont {Xu}}]{Bzdak:2019pkr}%
  \BibitemOpen
  \bibfield  {author} {\bibinfo {author} {\bibfnamefont {A.}~\bibnamefont {Bzdak}}, \bibinfo {author} {\bibfnamefont {S.}~\bibnamefont {Esumi}}, \bibinfo {author} {\bibfnamefont {V.}~\bibnamefont {Koch}}, \bibinfo {author} {\bibfnamefont {J.}~\bibnamefont {Liao}}, \bibinfo {author} {\bibfnamefont {M.}~\bibnamefont {Stephanov}},\ and\ \bibinfo {author} {\bibfnamefont {N.}~\bibnamefont {Xu}},\ }\bibfield  {title} {\bibinfo {title} {{Mapping the Phases of Quantum Chromodynamics with Beam Energy Scan}},\ }\href {https://doi.org/10.1016/j.physrep.2020.01.005} {\bibfield  {journal} {\bibinfo  {journal} {Phys. Rept.}\ }\textbf {\bibinfo {volume} {853}},\ \bibinfo {pages} {1} (\bibinfo {year} {2020})},\ \Eprint {https://arxiv.org/abs/1906.00936} {arXiv:1906.00936 [nucl-th]} \BibitemShut {NoStop}%
\bibitem [{\citenamefont {Steinheimer}\ and\ \citenamefont {Randrup}(2012)}]{Steinheimer:2012gc}%
  \BibitemOpen
  \bibfield  {author} {\bibinfo {author} {\bibfnamefont {J.}~\bibnamefont {Steinheimer}}\ and\ \bibinfo {author} {\bibfnamefont {J.}~\bibnamefont {Randrup}},\ }\bibfield  {title} {\bibinfo {title} {{Spinodal amplification of density fluctuations in fluid-dynamical simulations of relativistic nuclear collisions}},\ }\href {https://doi.org/10.1103/PhysRevLett.109.212301} {\bibfield  {journal} {\bibinfo  {journal} {Phys. Rev. Lett.}\ }\textbf {\bibinfo {volume} {109}},\ \bibinfo {pages} {212301} (\bibinfo {year} {2012})},\ \Eprint {https://arxiv.org/abs/1209.2462} {arXiv:1209.2462 [nucl-th]} \BibitemShut {NoStop}%
\bibitem [{\citenamefont {Savchuk}\ \emph {et~al.}(2023)\citenamefont {Savchuk}, \citenamefont {Poberezhnyuk}, \citenamefont {Motornenko}, \citenamefont {Steinheimer}, \citenamefont {Gorenstein},\ and\ \citenamefont {Vovchenko}}]{Savchuk:2022msa}%
  \BibitemOpen
  \bibfield  {author} {\bibinfo {author} {\bibfnamefont {O.}~\bibnamefont {Savchuk}}, \bibinfo {author} {\bibfnamefont {R.~V.}\ \bibnamefont {Poberezhnyuk}}, \bibinfo {author} {\bibfnamefont {A.}~\bibnamefont {Motornenko}}, \bibinfo {author} {\bibfnamefont {J.}~\bibnamefont {Steinheimer}}, \bibinfo {author} {\bibfnamefont {M.~I.}\ \bibnamefont {Gorenstein}},\ and\ \bibinfo {author} {\bibfnamefont {V.}~\bibnamefont {Vovchenko}},\ }\bibfield  {title} {\bibinfo {title} {{Phase transition amplification of proton number fluctuations in nuclear collisions from a transport model approach}},\ }\href {https://doi.org/10.1103/PhysRevC.107.024913} {\bibfield  {journal} {\bibinfo  {journal} {Phys. Rev. C}\ }\textbf {\bibinfo {volume} {107}},\ \bibinfo {pages} {024913} (\bibinfo {year} {2023})},\ \Eprint {https://arxiv.org/abs/2211.13200} {arXiv:2211.13200 [hep-ph]} \BibitemShut {NoStop}%
\bibitem [{\citenamefont {Vovchenko}\ \emph {et~al.}(2020)\citenamefont {Vovchenko}, \citenamefont {Savchuk}, \citenamefont {Poberezhnyuk}, \citenamefont {Gorenstein},\ and\ \citenamefont {Koch}}]{Vovchenko:2020tsr}%
  \BibitemOpen
  \bibfield  {author} {\bibinfo {author} {\bibfnamefont {V.}~\bibnamefont {Vovchenko}}, \bibinfo {author} {\bibfnamefont {O.}~\bibnamefont {Savchuk}}, \bibinfo {author} {\bibfnamefont {R.~V.}\ \bibnamefont {Poberezhnyuk}}, \bibinfo {author} {\bibfnamefont {M.~I.}\ \bibnamefont {Gorenstein}},\ and\ \bibinfo {author} {\bibfnamefont {V.}~\bibnamefont {Koch}},\ }\bibfield  {title} {\bibinfo {title} {{Connecting fluctuation measurements in heavy-ion collisions with the grand-canonical susceptibilities}},\ }\href {https://doi.org/10.1016/j.physletb.2020.135868} {\bibfield  {journal} {\bibinfo  {journal} {Phys. Lett. B}\ }\textbf {\bibinfo {volume} {811}},\ \bibinfo {pages} {135868} (\bibinfo {year} {2020})},\ \Eprint {https://arxiv.org/abs/2003.13905} {arXiv:2003.13905 [hep-ph]} \BibitemShut {NoStop}%
\bibitem [{\citenamefont {Sorensen}\ \emph {et~al.}(2023)\citenamefont {Sorensen}, \citenamefont {Oliinychenko}, \citenamefont {McLerran},\ and\ \citenamefont {Koch}}]{Sorensen:2022odd}%
  \BibitemOpen
  \bibfield  {author} {\bibinfo {author} {\bibfnamefont {A.}~\bibnamefont {Sorensen}}, \bibinfo {author} {\bibfnamefont {D.}~\bibnamefont {Oliinychenko}}, \bibinfo {author} {\bibfnamefont {L.}~\bibnamefont {McLerran}},\ and\ \bibinfo {author} {\bibfnamefont {V.}~\bibnamefont {Koch}},\ }\bibfield  {title} {\bibinfo {title} {{Measuring the Speed of Sound Using Cumulants of Baryon Number}},\ }\href {https://doi.org/10.5506/APhysPolBSupp.16.1-A48} {\bibfield  {journal} {\bibinfo  {journal} {Acta Phys. Polon. Supp.}\ }\textbf {\bibinfo {volume} {16}},\ \bibinfo {pages} {1} (\bibinfo {year} {2023})},\ \Eprint {https://arxiv.org/abs/2209.04957} {arXiv:2209.04957 [nucl-th]} \BibitemShut {NoStop}%
\bibitem [{\citenamefont {Gavin}\ \emph {et~al.}(2009)\citenamefont {Gavin}, \citenamefont {McLerran},\ and\ \citenamefont {Moschelli}}]{PhysRevC.79.051902}%
  \BibitemOpen
  \bibfield  {author} {\bibinfo {author} {\bibfnamefont {S.}~\bibnamefont {Gavin}}, \bibinfo {author} {\bibfnamefont {L.}~\bibnamefont {McLerran}},\ and\ \bibinfo {author} {\bibfnamefont {G.}~\bibnamefont {Moschelli}},\ }\bibfield  {title} {\bibinfo {title} {Long range correlations and the soft ridge in relativistic nuclear collisions},\ }\href {https://doi.org/10.1103/PhysRevC.79.051902} {\bibfield  {journal} {\bibinfo  {journal} {Phys. Rev. C}\ }\textbf {\bibinfo {volume} {79}},\ \bibinfo {pages} {051902} (\bibinfo {year} {2009})}\BibitemShut {NoStop}%
\bibitem [{\citenamefont {Dumitru}\ \emph {et~al.}(2008)\citenamefont {Dumitru}, \citenamefont {Gelis}, \citenamefont {McLerran},\ and\ \citenamefont {Venugopalan}}]{Dumitru:2008wn}%
  \BibitemOpen
  \bibfield  {author} {\bibinfo {author} {\bibfnamefont {A.}~\bibnamefont {Dumitru}}, \bibinfo {author} {\bibfnamefont {F.}~\bibnamefont {Gelis}}, \bibinfo {author} {\bibfnamefont {L.}~\bibnamefont {McLerran}},\ and\ \bibinfo {author} {\bibfnamefont {R.}~\bibnamefont {Venugopalan}},\ }\bibfield  {title} {\bibinfo {title} {{Glasma flux tubes and the near side ridge phenomenon at RHIC}},\ }\href {https://doi.org/10.1016/j.nuclphysa.2008.06.012} {\bibfield  {journal} {\bibinfo  {journal} {Nucl. Phys. A}\ }\textbf {\bibinfo {volume} {810}},\ \bibinfo {pages} {91} (\bibinfo {year} {2008})},\ \Eprint {https://arxiv.org/abs/0804.3858} {arXiv:0804.3858 [hep-ph]} \BibitemShut {NoStop}%
\bibitem [{\citenamefont {Kharzeev}\ and\ \citenamefont {Nardi}(2001)}]{Kharzeev:2000ph}%
  \BibitemOpen
  \bibfield  {author} {\bibinfo {author} {\bibfnamefont {D.}~\bibnamefont {Kharzeev}}\ and\ \bibinfo {author} {\bibfnamefont {M.}~\bibnamefont {Nardi}},\ }\bibfield  {title} {\bibinfo {title} {{Hadron production in nuclear collisions at RHIC and high density QCD}},\ }\href {https://doi.org/10.1016/S0370-2693(01)00457-9} {\bibfield  {journal} {\bibinfo  {journal} {Phys. Lett. B}\ }\textbf {\bibinfo {volume} {507}},\ \bibinfo {pages} {121} (\bibinfo {year} {2001})},\ \Eprint {https://arxiv.org/abs/nucl-th/0012025} {arXiv:nucl-th/0012025} \BibitemShut {NoStop}%
\bibitem [{\citenamefont {Shuryak}(2007)}]{PhysRevC.76.047901}%
  \BibitemOpen
  \bibfield  {author} {\bibinfo {author} {\bibfnamefont {E.}~\bibnamefont {Shuryak}},\ }\bibfield  {title} {\bibinfo {title} {Origin of the ``ridge'' phenomenon induced by jets in heavy ion collisions},\ }\href {https://doi.org/10.1103/PhysRevC.76.047901} {\bibfield  {journal} {\bibinfo  {journal} {Phys. Rev. C}\ }\textbf {\bibinfo {volume} {76}},\ \bibinfo {pages} {047901} (\bibinfo {year} {2007})}\BibitemShut {NoStop}%
\bibitem [{\citenamefont {Moschelli}\ and\ \citenamefont {Gavin}(2009)}]{Moschelli_2009}%
  \BibitemOpen
  \bibfield  {author} {\bibinfo {author} {\bibfnamefont {G.}~\bibnamefont {Moschelli}}\ and\ \bibinfo {author} {\bibfnamefont {S.}~\bibnamefont {Gavin}},\ }\bibfield  {title} {\bibinfo {title} {Two ridges, one explanation},\ }\href {https://doi.org/10.1016/j.nuclphysa.2009.10.058} {\bibfield  {journal} {\bibinfo  {journal} {Nuclear Physics A}\ }\textbf {\bibinfo {volume} {830}},\ \bibinfo {pages} {623c–624c} (\bibinfo {year} {2009})}\BibitemShut {NoStop}%
\bibitem [{\citenamefont {Gavin}\ and\ \citenamefont {Moschelli}(2008)}]{Gavin:2008ta}%
  \BibitemOpen
  \bibfield  {author} {\bibinfo {author} {\bibfnamefont {S.}~\bibnamefont {Gavin}}\ and\ \bibinfo {author} {\bibfnamefont {G.}~\bibnamefont {Moschelli}},\ }\bibfield  {title} {\bibinfo {title} {{Viscosity and the Soft Ridge at RHIC}},\ }\href {https://doi.org/10.1088/0954-3899/35/10/104084} {\bibfield  {journal} {\bibinfo  {journal} {J. Phys. G}\ }\textbf {\bibinfo {volume} {35}},\ \bibinfo {pages} {104084} (\bibinfo {year} {2008})},\ \Eprint {https://arxiv.org/abs/0806.4366} {arXiv:0806.4366 [nucl-th]} \BibitemShut {NoStop}%
\bibitem [{\citenamefont {Gavin}\ and\ \citenamefont {Abdel-Aziz}(2006)}]{PhysRevLett.97.162302}%
  \BibitemOpen
  \bibfield  {author} {\bibinfo {author} {\bibfnamefont {S.}~\bibnamefont {Gavin}}\ and\ \bibinfo {author} {\bibfnamefont {M.}~\bibnamefont {Abdel-Aziz}},\ }\bibfield  {title} {\bibinfo {title} {Measuring shear viscosity using transverse momentum correlations in relativistic nuclear collisions},\ }\href {https://doi.org/10.1103/PhysRevLett.97.162302} {\bibfield  {journal} {\bibinfo  {journal} {Phys. Rev. Lett.}\ }\textbf {\bibinfo {volume} {97}},\ \bibinfo {pages} {162302} (\bibinfo {year} {2006})}\BibitemShut {NoStop}%
\bibitem [{\citenamefont {Savchuk}\ and\ \citenamefont {Pratt}(2024)}]{first_paper}%
  \BibitemOpen
  \bibfield  {author} {\bibinfo {author} {\bibfnamefont {O.}~\bibnamefont {Savchuk}}\ and\ \bibinfo {author} {\bibfnamefont {S.}~\bibnamefont {Pratt}},\ }\bibfield  {title} {\bibinfo {title} {{Correlations of conserved quantities at finite baryon density}},\ }\href {https://doi.org/10.1103/PhysRevC.109.024910} {\bibfield  {journal} {\bibinfo  {journal} {Phys. Rev. C}\ }\textbf {\bibinfo {volume} {109}},\ \bibinfo {pages} {024910} (\bibinfo {year} {2024})},\ \Eprint {https://arxiv.org/abs/2311.02046} {arXiv:2311.02046 [nucl-th]} \BibitemShut {NoStop}%
\bibitem [{\citenamefont {Adams}\ \emph {et~al.}(2005)\citenamefont {Adams} \emph {et~al.}}]{PhysRevC.72.014904}%
  \BibitemOpen
  \bibfield  {author} {\bibinfo {author} {\bibfnamefont {J.}~\bibnamefont {Adams}} \emph {et~al.} (\bibinfo {collaboration} {STAR Collaboration and STAR-RICH Collaboration}),\ }\bibfield  {title} {\bibinfo {title} {Azimuthal anisotropy in au+au collisions at $\sqrt{{s}_{\mathit{NN}}}=200\phantom{\rule{0.3em}{0ex}}\mathrm{GeV}$},\ }\href {https://doi.org/10.1103/PhysRevC.72.014904} {\bibfield  {journal} {\bibinfo  {journal} {Phys. Rev. C}\ }\textbf {\bibinfo {volume} {72}},\ \bibinfo {pages} {014904} (\bibinfo {year} {2005})}\BibitemShut {NoStop}%
\bibitem [{\citenamefont {Heinz}\ and\ \citenamefont {Snellings}(2013)}]{Heinz_2013}%
  \BibitemOpen
  \bibfield  {author} {\bibinfo {author} {\bibfnamefont {U.}~\bibnamefont {Heinz}}\ and\ \bibinfo {author} {\bibfnamefont {R.}~\bibnamefont {Snellings}},\ }\bibfield  {title} {\bibinfo {title} {Collective flow and viscosity in relativistic heavy-ion collisions},\ }\href {https://doi.org/10.1146/annurev-nucl-102212-170540} {\bibfield  {journal} {\bibinfo  {journal} {Annual Review of Nuclear and Particle Science}\ }\textbf {\bibinfo {volume} {63}},\ \bibinfo {pages} {123–151} (\bibinfo {year} {2013})}\BibitemShut {NoStop}%
\bibitem [{\citenamefont {Pratt}\ and\ \citenamefont {Plumberg}(2021)}]{PhysRevC.104.014906}%
  \BibitemOpen
  \bibfield  {author} {\bibinfo {author} {\bibfnamefont {S.}~\bibnamefont {Pratt}}\ and\ \bibinfo {author} {\bibfnamefont {C.}~\bibnamefont {Plumberg}},\ }\bibfield  {title} {\bibinfo {title} {Charge balance functions for heavy-ion collisions at energies available at the cern large hadron collider},\ }\href {https://doi.org/10.1103/PhysRevC.104.014906} {\bibfield  {journal} {\bibinfo  {journal} {Phys. Rev. C}\ }\textbf {\bibinfo {volume} {104}},\ \bibinfo {pages} {014906} (\bibinfo {year} {2021})}\BibitemShut {NoStop}%
\bibitem [{\citenamefont {Agakishiev}\ \emph {et~al.}(2011)\citenamefont {Agakishiev} \emph {et~al.}}]{STAR:2011iio}%
  \BibitemOpen
  \bibfield  {author} {\bibinfo {author} {\bibfnamefont {H.}~\bibnamefont {Agakishiev}} \emph {et~al.} (\bibinfo {collaboration} {STAR}),\ }\bibfield  {title} {\bibinfo {title} {{Evolution of the differential transverse momentum correlation function with centrality in Au$+$Au collisions at $\sqrt{s_{NN}} =$ 200 GeV}},\ }\href {https://doi.org/10.1016/j.physletb.2011.09.075} {\bibfield  {journal} {\bibinfo  {journal} {Phys. Lett. B}\ }\textbf {\bibinfo {volume} {704}},\ \bibinfo {pages} {467} (\bibinfo {year} {2011})},\ \Eprint {https://arxiv.org/abs/1106.4334} {arXiv:1106.4334 [nucl-ex]} \BibitemShut {NoStop}%
\bibitem [{\citenamefont {Romatschke}(2010)}]{Romatschke:2010}%
  \BibitemOpen
  \bibfield  {author} {\bibinfo {author} {\bibfnamefont {P.}~\bibnamefont {Romatschke}},\ }\bibfield  {title} {\bibinfo {title} {{New Developments in Relativistic Viscous Hydrodynamics}},\ }\href {https://doi.org/10.1142/S0218301310014613} {\bibfield  {journal} {\bibinfo  {journal} {Int. J. Mod. Phys. E}\ }\textbf {\bibinfo {volume} {19}},\ \bibinfo {pages} {1} (\bibinfo {year} {2010})},\ \Eprint {https://arxiv.org/abs/0902.3663} {arXiv:0902.3663 [hep-ph]} \BibitemShut {NoStop}%
\bibitem [{\citenamefont {Bazavov}\ \emph {et~al.}(2014)\citenamefont {Bazavov} \emph {et~al.}}]{Bazavov_2014}%
  \BibitemOpen
  \bibfield  {author} {\bibinfo {author} {\bibfnamefont {A.}~\bibnamefont {Bazavov}} \emph {et~al.},\ }\bibfield  {title} {\bibinfo {title} {Equation of state in (2)-flavor qcd},\ }\bibfield  {journal} {\bibinfo  {journal} {Physical Review D}\ }\textbf {\bibinfo {volume} {90}},\ \href {https://doi.org/10.1103/physrevd.90.094503} {10.1103/physrevd.90.094503} (\bibinfo {year} {2014})\BibitemShut {NoStop}%
\bibitem [{\citenamefont {Bazavov}\ \emph {et~al.}(2012)\citenamefont {Bazavov} \emph {et~al.}}]{Bazavov_2012}%
  \BibitemOpen
  \bibfield  {author} {\bibinfo {author} {\bibfnamefont {A.}~\bibnamefont {Bazavov}} \emph {et~al.},\ }\bibfield  {title} {\bibinfo {title} {Fluctuations and correlations of net baryon number, electric charge, and strangeness: A comparison of lattice qcd results with the hadron resonance gas model},\ }\bibfield  {journal} {\bibinfo  {journal} {Physical Review D}\ }\textbf {\bibinfo {volume} {86}},\ \href {https://doi.org/10.1103/physrevd.86.034509} {10.1103/physrevd.86.034509} (\bibinfo {year} {2012})\BibitemShut {NoStop}%
\bibitem [{\citenamefont {Huovinen}\ \emph {et~al.}(2017)\citenamefont {Huovinen}, \citenamefont {Lo}, \citenamefont {Marczenko}, \citenamefont {Morita}, \citenamefont {Redlich},\ and\ \citenamefont {Sasaki}}]{Huovinen:2016xxq}%
  \BibitemOpen
  \bibfield  {author} {\bibinfo {author} {\bibfnamefont {P.}~\bibnamefont {Huovinen}}, \bibinfo {author} {\bibfnamefont {P.~M.}\ \bibnamefont {Lo}}, \bibinfo {author} {\bibfnamefont {M.}~\bibnamefont {Marczenko}}, \bibinfo {author} {\bibfnamefont {K.}~\bibnamefont {Morita}}, \bibinfo {author} {\bibfnamefont {K.}~\bibnamefont {Redlich}},\ and\ \bibinfo {author} {\bibfnamefont {C.}~\bibnamefont {Sasaki}},\ }\bibfield  {title} {\bibinfo {title} {{Effects of \ensuremath{\rho}-meson width on pion distributions in heavy-ion collisions}},\ }\href {https://doi.org/10.1016/j.physletb.2017.03.060} {\bibfield  {journal} {\bibinfo  {journal} {Phys. Lett. B}\ }\textbf {\bibinfo {volume} {769}},\ \bibinfo {pages} {509} (\bibinfo {year} {2017})},\ \Eprint {https://arxiv.org/abs/1608.06817} {arXiv:1608.06817 [hep-ph]} \BibitemShut {NoStop}%
\bibitem [{\citenamefont {Pratt}\ \emph {et~al.}(2018)\citenamefont {Pratt}, \citenamefont {Kim},\ and\ \citenamefont {Plumberg}}]{Pratt:2017oyf}%
  \BibitemOpen
  \bibfield  {author} {\bibinfo {author} {\bibfnamefont {S.}~\bibnamefont {Pratt}}, \bibinfo {author} {\bibfnamefont {J.}~\bibnamefont {Kim}},\ and\ \bibinfo {author} {\bibfnamefont {C.}~\bibnamefont {Plumberg}},\ }\bibfield  {title} {\bibinfo {title} {{Evolution of Charge Fluctuations and Correlations in the Hydrodynamic Stage of Heavy Ion Collisions}},\ }\href {https://doi.org/10.1103/PhysRevC.98.014904} {\bibfield  {journal} {\bibinfo  {journal} {Phys. Rev. C}\ }\textbf {\bibinfo {volume} {98}},\ \bibinfo {pages} {014904} (\bibinfo {year} {2018})},\ \Eprint {https://arxiv.org/abs/1712.09298} {arXiv:1712.09298 [nucl-th]} \BibitemShut {NoStop}%
\bibitem [{\citenamefont {Pradeep}\ \emph {et~al.}(2022)\citenamefont {Pradeep}, \citenamefont {Rajagopal}, \citenamefont {Stephanov},\ and\ \citenamefont {Yin}}]{Pradeep:2022mkf}%
  \BibitemOpen
  \bibfield  {author} {\bibinfo {author} {\bibfnamefont {M.}~\bibnamefont {Pradeep}}, \bibinfo {author} {\bibfnamefont {K.}~\bibnamefont {Rajagopal}}, \bibinfo {author} {\bibfnamefont {M.}~\bibnamefont {Stephanov}},\ and\ \bibinfo {author} {\bibfnamefont {Y.}~\bibnamefont {Yin}},\ }\bibfield  {title} {\bibinfo {title} {{Freezing out fluctuations in Hydro+ near the QCD critical point}},\ }\href {https://doi.org/10.1103/PhysRevD.106.036017} {\bibfield  {journal} {\bibinfo  {journal} {Phys. Rev. D}\ }\textbf {\bibinfo {volume} {106}},\ \bibinfo {pages} {036017} (\bibinfo {year} {2022})},\ \Eprint {https://arxiv.org/abs/2204.00639} {arXiv:2204.00639 [hep-ph]} \BibitemShut {NoStop}%
\bibitem [{\citenamefont {Abelev}\ \emph {et~al.}(2013)\citenamefont {Abelev} \emph {et~al.}}]{PhysRevC.88.044910}%
  \BibitemOpen
  \bibfield  {author} {\bibinfo {author} {\bibfnamefont {B.}~\bibnamefont {Abelev}} \emph {et~al.} (\bibinfo {collaboration} {ALICE Collaboration}),\ }\bibfield  {title} {\bibinfo {title} {Centrality dependence of $\ensuremath{\pi}$, $k$, and $p$ production in pb-pb collisions at $\sqrt{{s}_{NN}}=2.76$ tev},\ }\href {https://doi.org/10.1103/PhysRevC.88.044910} {\bibfield  {journal} {\bibinfo  {journal} {Phys. Rev. C}\ }\textbf {\bibinfo {volume} {88}},\ \bibinfo {pages} {044910} (\bibinfo {year} {2013})}\BibitemShut {NoStop}%
\end{thebibliography}%

\end{document}